\documentclass[english,twocolumn,prl,showpacs]{revtex4}
\usepackage[T1]{fontenc}
\usepackage[latin1]{inputenc}
\usepackage{graphicx}

\makeatletter

\usepackage{amsthm}

\usepackage{epsfig}

\makeatletter

\makeatletter

\makeatletter

\usepackage{epsfig}
\usepackage{amsthm}

\makeatother

\makeatother

\usepackage{babel}
\makeatother
\begin{document}

\title{Universal shape law of stochastic supercritical bifurcations:
Theory and experiments.}

\author{Gonzague Agez, Marcel G. Clerc, and Eric Louvergneaux$^{\ddag}$}

\affiliation{Departamento de F\'{\i}sica, Facultad de Ciencias F\'{\i}sicas
y Matem\'{a}ticas, Universidad de Chile, Casilla 487-3, Santiago,
Chile,}

\affiliation{$^{\ddag}$Laboratoire de Physique des Lasers, Atomes
et Mol\'{e}cules, UMR-CNRS 8523, CERLA FR-CNRS 2416,
Universit\'{e} des Sciences et Technologies de Lille, 59655
Villeneuve d'Ascq Cedex, France.}

\begin{abstract}
A universal law for the supercritical bifurcation shape of
transverse one-dimensional (1D) systems in presence of additive
noise is given. The stochastic Langevin equation of such systems
is solved by using a Fokker-Planck equation leading to the
expression for the  most probable amplitude of the critical mode.
From this universal expression, the shape of the bifurcation, its
location and its evolution with the noise level are completely
defined. Experimental results obtained for a 1D transverse
Kerr-like slice subjected to optical feedback are in excellent
agreement.
\end{abstract}


\maketitle In nature, most of physical systems are subjected to
fluctuations. For a long time, the effects of these fluctuations
were either considered as a nuisance (degradation of the
signal-to-noise ratio) or ignored because it  was not known how to
handle them. Since three decades, a wealth of theoretical and
experimental researches have shown that fluctuations can have
rather surprisingly constructive and counterintuitive effects in
many physical systems and that they can be figured out with the
help of different analysis tools. These situations occur when
there are mechanisms of noise amplification or when noise
interacts with nonlinearities or driving forces on the system. The
most well-know examples in zero dimensional systems are noise
induced transition \cite{HL84} and stochastic resonance
\cite{GHJM98}. More recently, examples in spatially extended
system are noise induced phase transition,  noise-induced patterns
(see \cite{OS99} and references therein), noise-sustained
structures in convective instability \cite{AGTL06}, stochastic
spatio-temporal intermittency \cite{SanMiguiel}, noise-induced
travelling waves \cite{Zhou}, noise-induced ordering transition
\cite{MLKB97} and front propagation \cite{ClercNoise2006}. A
direct consequence of noise effects is the modification of the
deterministic bifurcation shapes. The critical points and the
physical mechanisms are masked by fluctuations. It is important to
remark that the critical points generically represent a change of
balance between forces. Hence, the characterization of noisy
bifurcations is a fundamental problem due to the ubiquitous nature
of bifurcations. For instance, the supercritical bifurcations
transform into smooth transitions between the two states and the
subcritical bifurcations experience hysteresis size modifications.
In absence of noise, the shape of a bifurcation and its
characteristics are given by the analytical solution of the
deterministic amplitude equation of the critical mode
\cite{CrossHohenberg}. On the other hand, in presence of noise, no
such analytical expression can be obtained from the stochastic
amplitude equation. In this latter situation, the below and above
bifurcation point regimes can be treated separately, but without
continuity between their respective solutions. For instance, in
noisy spatially extended systems in which the systems are
characterized by the appearance of pattern precursors below the
bifurcation point and by established patterns that fluctuate above
this point \cite{Gonzague}, the precursor amplitude \cite{SZ93},
obtained from the linear study of the stochastic equation,
diverges at the bifurcation point and does not connect to the
{}``mean'' amplitude of the fluctuating pattern, obtained from the
deterministic equation. To our knowledge, no universal analytical
expression of the critical mode amplitude, describing the complete
transition from below to above the bifurcation point, exists for
the supercritical bifurcations in presence of noise.

In this letter, we propose a universal description of the
supercritical bifurcation shapes of 1D transverse systems (either
uniforms or very slowly varying in space) in presence of noise
that is also valid for the second order bifurcations of temporal
(zero-dimensional) systems. \emph{More precisely, we give a
unified expression for the most probable amplitude describing the
supercritical bifurcations in presence of noise, including the
noise level and the bifurcation point location.} The systems under
study are described by stochastic partial differential equations
of the Langevin type \cite{VanKampen} (first order in time and
with linear noise terms) involving additive white noise. Firstly,
we solve the Langevin equation describing the stochastic dynamics
by using a Fokker-Planck equation for the probability density of
the critical mode amplitude. Then, from the stationary
distribution of this amplitude, we deduce the bifurcation shape by
means of the most probable value of the pattern amplitude.
Finally, the comparison with experimental results obtained in  a
Kerr-like slice subjected to 1D optical feedback is given and
leads to an excellent agreement.

Let us consider a 1D extended system that exhibits a
supercritical spatial bifurcation described by \begin{equation}
\partial_{t}\vec{u}=\vec{f}(\vec{u},\partial_{x},\left\{ \mu\right\} )
+\sqrt{\eta}\vec{\zeta}(x,t)\label{E-pde}\end{equation}
 where $\vec{u}\left(x,t\right)$ is a field that describes the system
under study, $\vec{f}$ is the vector field, $\left\{ \mu\right\} $
is a set of parameters that characterizes the system, $\eta$ is
the noise level intensity, and $\vec{\zeta}(x,t)$ is a white
Gaussian noise with zero mean value and correlation $\left\langle
\zeta^{i}(x,t) \zeta^{j}(x{\acute{}},t{\acute{}})\right\rangle
=\delta^{ij}\delta
\left(t{\acute{}}-t\right)\delta\left(x{\acute{}}-x\right)$.

\begin{figure}[t]

\begin{centering}\includegraphics{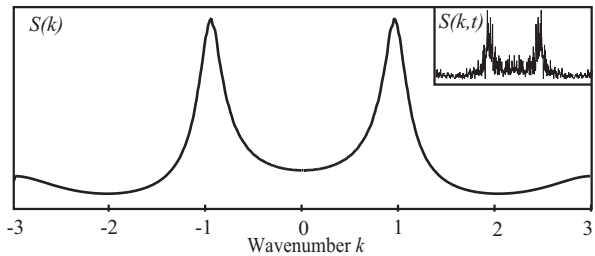}\par\end{centering}

\caption{Averaged Fourier transform modulus $S(k)$ of the field
$\vec{u}$ for the supercritical Swift-Hohenberg model \cite{OS99}
obtained below the bifurcation point. The inset figure corresponds
to an instantaneous snapshot of $S(k,t)$.
\label{F-StructureFactor}}
\end{figure}

We assume that the associated deterministic system  ($\eta=0$)
possesses a stationary state $\vec{u}_{0}$, that satisfies
$\vec{f} (\vec{u}_{o},\partial_{x},\left\{ \mu\right\} )=0$. Above
a set of critical values $\left\{ \mu_{c}\right\} $, the system
exhibits a spatial instability such that $\vec{u}$ reads $\vec{u}
\left(x,t\right)=\vec{u}_{0}+e^{\lambda(k)t+ikx}\hat{u},$ where
$\lambda\left(k\right)$ is the linear growth rate and $k$ the wave
number of the instability. Close to $\left\{ \mu_{c}\right\} $,
for $\lambda\left(k\right)<0$ (i.e. $\vec{u}_{0}$ is stable), the
profile of $\lambda$ already displays a maximum for a non null
wave number close to the critical one $k_{c}$. Hence, below the
bifurcation point, when noise is present, among all the excited
spatial modes, the one associated with the maximum growth rate
($k\simeq k_{c}$) rules the dynamics. The dynamical behavior of
the system is then characterized by pattern precursors. Figure
\ref{F-StructureFactor} illustrates this property. It shows the
instantaneous $S\left(k,t\right)$ and the averaged
$S\left(k\right)=<S(k,t)>_{t}$ modulus of the Fourier transform of
the field $\vec{u}$, also called the structure factor, for the
supercritical Swift-Hohenberg equation \cite{OS99} below the
deterministic threshold. We can remark that the maxima of the
function, both instantaneous and averaged, already give the
incoming critical wave numbers $k_{c}=\pm1$
(Fig.\ref{F-StructureFactor}).

\begin{figure}[b]

\noindent \begin{centering}\includegraphics{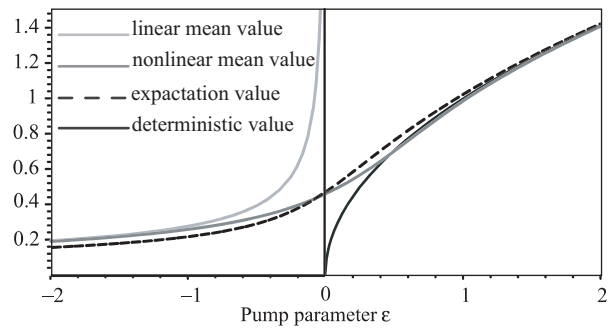}\par\end{centering}

\caption{Bifurcation diagrams of the amplitude of the critical
mode obtained for different solutions of Eq.
(\ref{E-StochasticAmplitudeEquation}). Namely, the deterministic
value $\emph{A}$, the expectation value $|a_{max}|$, the linear
and non-linear mean values $\langle|a|\rangle_{linear}$ and
$\langle|a|\rangle_{nonlinear}$ respectively (see text for value
definitions).}

\label{F-BifurcationDiagram}
\end{figure}

To capture the dynamics of the system of Eq. (\ref{E-pde})  in an
unified description close to the instability threshold, we
introduce the ansatz
$\vec{u}\left(x,t\right)=\vec{u}_{o}+A\left(T=\varepsilon
t,X=\sqrt{\varepsilon}x\right)e^{ik_{c}x}\hat{u}_{k}+\bar{A}
\left(T,X\right)e^{-ik_{c}x}\hat{u}_{k}+h.o.t$
\cite{CrossHohenberg}, where $A$ is the amplitude of critical
mode, $\varepsilon$ is the bifurcation parameter which is
proportional to $\left\Vert \mu-\mu_{c}\right\Vert $,
$\hat{u}_{k}$ is the eigenvector associated to the critical mode,
$\left\{ T,X\right\} $ are the slow time and spatial  variables
and the high order terms ($h.o.t$) are a series in the amplitude
$A$. Introducing the above ansatz in Eq. (\ref{E-pde})  leads to
the amplitude equation that satisfies the solvability condition
\cite{CrossHohenberg}

\begin{equation}
\partial_{T}A=\varepsilon A-|A|^{2}A+\partial_{XX}A+\sqrt{\eta}
\xi\left(X,T\right),
\label{E-StochasticAmplitudeEquation}
\end{equation}
 where $\xi\left(X,T\right)$ is a white Gaussian noise  with null
mean value and correlation $\left\langle \xi\left(X,T\right)\xi
\left(X^{\prime},T^{\prime}\right)\right\rangle =0$
and $\left\langle \xi\left(X,T\right)\bar{\xi}\left(X^{\prime},
T^{\prime}\right)\right\rangle =\delta\left(T{\acute{}}-T\right)
\delta\left(X{\acute{}}-X\right)$.
The $\xi\left(X,T\right)$ term is then a linear combination of the
original noise source term $\vec{\zeta}(x,t)$. In absence of noise
($\eta=0$), when the bifurcation parameter $\mathcal{\varepsilon}$
is negative, the system is characterized by a stable equilibrium state
of amplitude $A=0$, and when $\mathcal{\varepsilon}$ becomes positive,
the system exhibits a degenerate pitchfork bifurcation whose amplitude
evolves as $|A|\sim\sqrt{\varepsilon}$ (deterministic value in Fig.
\ref{F-BifurcationDiagram}).

\begin{figure}[t]
 \includegraphics{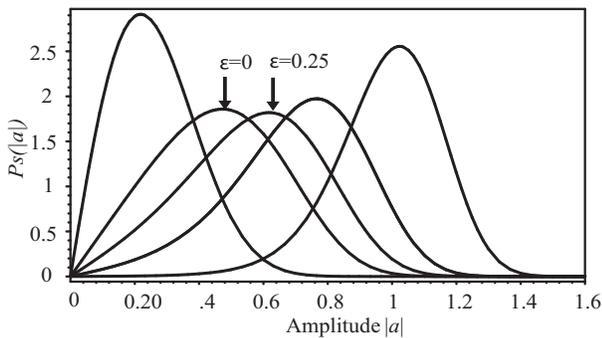}

\caption{Stationary probability distribution $P_{s}\left(|a|\right)$ for
different values of the bifurcation parameter $\varepsilon$. From
left to right: $\varepsilon=-1;0;0.25;0.5;1$. $\eta=0.2$.}

\label{F-Probability}
\end{figure}

The next step is to consider the global amplitude
$a\left(T\right)=\int_{-L/2}^{L/2}A(T,X)dX/L,$ where $L$ is the
size of the transverse domain under study. This quantity is
relevant to describe the bifurcation as soon as the transverse
domain is either uniform or with very slow spatial variations of
the control parameters. Experimentally, it will be achieved by
choosing a restricted part $L$ of the transverse pattern within
which the control parameters are slowly varying. This restriction
excludes all specific cases with fast spatial variations (e.g.
local structures), which require specific analyses associated with
the considered transverse profiles, but they are out of the scope
of this paper. The global amplitude $a\left(T\right)$ then
satisfies the Langevin equation
\begin{equation}
\partial_{T}a=\varepsilon a-|a|^{2}a+\sqrt{\eta}\varsigma\left(T\right),
\label{E-Langevine}\end{equation}
 where $\varsigma\left(T\right)\equiv\int_{-L/2}^{L/2}\xi(T,X)dX/L$
is a Gaussian white noise with zero expectation value and
correlation $\left\langle
\varsigma\left(T\right)\bar{\varsigma}\left(T^{\prime}
\right)\right\rangle =\delta\left(T{\acute{}}-T\right).$ The
general way of obtaining a solution of such Langevin  equation is
by use of a Fokker-Planck equation which provides us with a
deterministic equation satisfied by the time dependent probability
density $P(a,\bar{a};T)$ of the amplitude $a$ \cite{VanKampen}
that reads

\begin{equation}
\partial_{T}P=\partial_{a}\left\{ -\varepsilon a+|a|^{2}a+
\frac{\eta}{2}\partial_{\bar{a}}\right\} P+c.c.
\label{eq:E-FokkerPlanck}
\end{equation}
where $c.c.$ means \char`\"{}complex conjugate\char`\"{}. The
associated stationary probability density of the modulus of
$\emph{a}$ is

\begin{equation}
P_{s}\left(|a|,\varepsilon,\eta\right)= Q(\varepsilon,\eta) |a|
e^{\frac{\varepsilon|a|^{2}-\frac{|a|^{4}}{2}}{\eta}},
\label{E-StationaryProbability}
\end{equation}
where $Q(\varepsilon,\eta)\equiv 2\sqrt{2}
e^{-\frac{\varepsilon^{2}} {2\eta}} /
erfc(-\frac{\varepsilon}{\sqrt{2\eta}}) \sqrt{\pi\eta}$. The
stationary probability density is shown in Fig.
\ref{F-Probability} for different values of the bifurcation
parameter $\varepsilon$. The probability density function is not
symmetrical with respect to its maximum so that the most relevant
quantity for characterizing
$P_{s}\left(|a|,\varepsilon,\eta\right)$ is its maximum and not
its mean value as usually calculated e.g. in experiments. The
value of $|a|$ corresponding to  the maximum of
$P_{s}\left(|a|,\varepsilon,\eta\right)$ occurs at the expectation
value $|a_{max}|$ given by \begin{equation}
|a_{max}|=\sqrt{\frac{\varepsilon+\sqrt{\varepsilon^{2}+2\eta}}{2}}
\label{E-expectationvalue}\end{equation}

\begin{figure}[t]
 \includegraphics{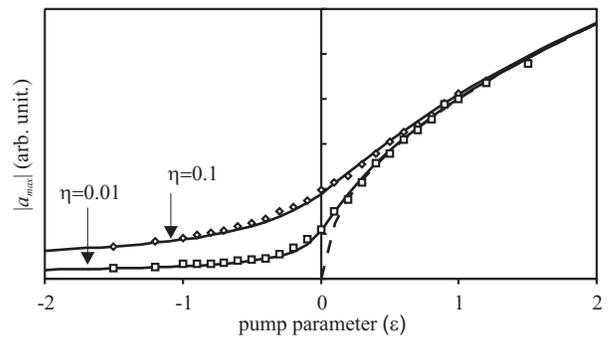}

\caption{Influence of the noise level $\eta$ on the shape of the
unperfect supercritical bifurcation. The continuous curves
correspond to the analytical expectation value $|a_{max}|$ of Eq.
(\ref{E-expectationvalue}). The squares and losanges are obtained
from the numerical simulations of the stochastic Swift-Hohenberg
equation Eq. (\ref{E-Langevine}). \label{F-PrecursorSH}}
\end{figure}

\noindent When the bifurcation parameter $\varepsilon$ is driven
far from 0 ($|\varepsilon|\gg1$), this expectation value converge
to zero ($\sqrt{-\eta/2\varepsilon}$) for negative  values of
$\varepsilon$ and to $\sqrt{\varepsilon}$ (as for the
deterministic case) for positive values. Plotting this quantity
versus $\varepsilon$ gives the supercritical bifurcation shape in
presence of noise (dashed line in Fig.
\ref{F-BifurcationDiagram}). Let's now discuss the relevance of
$|a_{max}|$ for describing the noisy supercritical bifurcation. If
we neglect the nonlinear term in the Langevin equation
(\ref{E-Langevine}), one can perform a linear stability analysis
that provides us with the linear mean value of the amplitude
modulus $\left\langle |a|\right\rangle
_{linear}=\sqrt{-\pi\eta/4\varepsilon}$. Note that this value
diverges at the bifurcation point and is only valid for
$\varepsilon<0$ (Fig. \ref{F-BifurcationDiagram}). The
corresponding nonlinear mean value $\left\langle |a|\right\rangle
_{nonlinear}$ is computed numerically from the probability density
$P_{s}(|a|)$. All the linear mean, non-linear mean, expected and
deterministic values of the critical mode amplitude are reported
on Fig. \ref{F-BifurcationDiagram} for comparison. The interesting
region is located in the vicinity of the bifurcation point where
the behaviors of the different curves strongly differ. The linear
mean value and the deterministic value do not correspond to a
realistic physical behavior since the amplitude never diverges at
threshold and we are considering a noisy system respectively. Only
the nonlinear mean and expected values can mimic the supercritical
bifurcation in presence of noise. However, as we have mentioned
earlier, due to the asymmetry of
$P_{s}\left(|a|,\varepsilon,\eta\right)$, the most relevant value
for describing the evolution of the amplitude versus the control
parameter is the expectation value $|a_{max}|$. As this value
includes the noise level $\eta$, we have performed numerical
simulations including noise to study the influence of noise on the
bifurcation shape. Figure \ref{F-PrecursorSH} clearly depicts the
continuous deformation and evolution of the bifurcation shape with
the level of noise. It also shows the very good agreement between
the numerical values of $|a_{max}|$ obtained from the numerical
simulations of the stochastic supercritical Swift-Hohenberg
equation \cite{SwiftHohenberg} and its analytical value {[}Eq.
(\ref{E-expectationvalue})]. Thus, the expression of $|a_{max}|$
is the relevant quantity for describing the noisy supercritical
bifurcations including the noise level and the bifurcation point
location ($\varepsilon=0$) of systems satisfying the Eq.
(\ref{E-pde}).

\noindent Regardless of the sign of $\varepsilon$,  the width of
the stationary probability distribution decreases as
$|\varepsilon|$ increases far from 0 (Fig. \ref{F-Probability}).
Therefore, close to the bifurcation point ($\varepsilon=0$) the
value of $P_{s}\left(|a_{max}|\right)$ exhibits a minimum versus
$\varepsilon$ ($\varepsilon\approx0.25$ in the case of Fig.
\ref{F-Probability}) which is shifted from this bifurcation point.
The width of the stationary probability density is then maximum
corresponding to a dynamics characterized by large amplitude
fluctuations. This shift between the minimum of
$P_{s}\left(|a_{max}|\right)$ and the bifurcation point
($\varepsilon=0$) reminds the bifurcation postponements as in
\cite{LT86}. After a straightforward calculation, the value of the
bifurcation parameter ($\varepsilon_{min}$) associated with the
minimum of $P_{s}\left(|a_{max}|\right)$, for a given value of
noise intensity, is \[
\varepsilon_{min}\left(\eta\right)\approx0.55\sqrt{\eta}.\]

\noindent It allow us to propose a criterion for the localization
of the bifurcation point: \textit{\emph{for a given noise intensity
(}}\emph{$\eta$}\textit{\emph{) the bifurcation point is localized
at}} \emph{$\varepsilon_{min}-0.55\sqrt{\eta}$}\textit{\emph{, where}}
\emph{$\varepsilon_{min}$} \textit{\emph{is the value of the control
parameter for which}} \emph{$P_{s}\left(|a_{max}|,\varepsilon\right)$}
\textit{\emph{is minimum.}}

\begin{figure}[t]
 \includegraphics{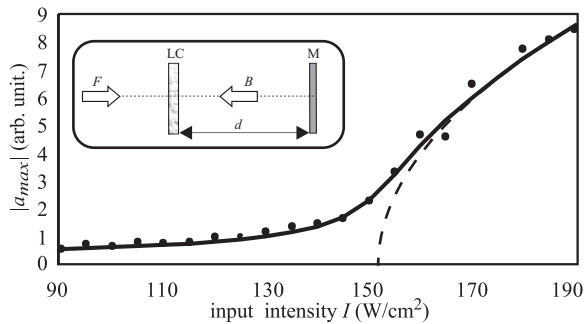}
\caption{Experimental bifurcation diagram for the optical feedback system.
The dots are the experimental measured values of $|a_{max}|$ and
the continuous curve its fitted analytical value (Eq. \ref{E-expectationvalue}).
The dashed line is the corresponding deterministic bifurcation deduced
from the fit. The inset figure is a schematic sketch of the experimental
set-up. LC liquid crystal layer; M feedback mirror; $F$ input optical
field; $B$ backward optical field; $d$ feedback length. }

\label{F-experiment}
\end{figure}

In order to completely validate the universal law of Eq.
(\ref{E-expectationvalue}), we have applied our analysis to
experiments realized on a noisy 1D transverse system known to
exhibit a supercritical bifurcation at the onset of roll pattern
formation. The system is a nematic liquid crystal slice subjected
to optical feedback (cf. Fig. \ref{F-experiment}) based on the
well known feedback optical system \cite{akhman88,firth90}. The
corresponding stochastic model reads \cite{AGTL06}
\begin{equation}
\partial_{t}\vec{u}=\left(\partial_{xx}-1\right)\vec{u}+
\left|F\right|^{2}+R\left|e^{i\sigma\partial_{xx}}\left(e^{i\chi\vec{u}}
F\right)\right|^{2}+\sqrt{\eta}\vec{\zeta}\label{eq:CL}\end{equation}
 where $\vec{u}(x,t)$ stands for the refractive index of the nonlinear
nematic liquid crystal (LC) layer; $t$ and $x$ are the time and
transverse space variables scaled with respect to the relaxation
time $\tau$ and the diffusion length $l_{d}$; $R$ is the mirror
intensity reflectivity. $\sigma=d/k_{0}$ where $d$ is the
slice-mirror distance and $k_{0}$ is the optical wave number of the
field. $F$ is the forward input optical field, its transverse
profile is accounted for using $F(x)=F_{0}\exp(-x^{2}/w^{2})$ for a
Gaussian pump beam of radius $w$. $\zeta$ and $\eta$ are the noise
source and level respectively as defined in Eq. (\ref{E-pde}). The
Kerr effect is parametrized by $\chi$ which is positive (negative)
for a focusing (defocusing) medium.

Eq. (\ref{eq:CL}) is similar to Eq. (\ref{E-pde}) and the spatial
variations of the pumping beam around its maximum are slow (less
than 5\% for a domain width $L$ including 4 to 5 rolls) due to the
high transverse aspect ratio ($2wk_c/2\pi\geq30$), so that the
previous analysis can be applied to describe the supercritical
bifurcation of our experimental noisy system. On Fig.
\ref{F-experiment} we have plotted the experimental recordings of
the amplitude expectation value $|a_{max}|$ together with its
analytical expression {[}Eq. (\ref{E-expectationvalue})] for a
transverse domain width $L$ including 4 to 5 rolls around the
center of the Gaussian pumping beam. We can see that the
analytical expression fits very well the experimental values. It
provides us with the deterministic threshold and noise level,
$I_{th}=\left|F_{th}\right|^{2}=152\: W.cm^{-2}$ and $\eta=0.01$
respectively. $\eta$ is in very good agreement with previous
determinations \cite{Gonzague,Gonzague2}. So, the universal
amplitude expression of Eq. (\ref{E-expectationvalue}) is valid
and relevant to describe the supercritical spatial bifurcation
shape of our noisy system even and more generally to describe the
supercritical bifurcations of 1D systems in presence of noise .

In conclusion, we have given an universal amplitude equation for 1D
systems in presence of noise. From this equation, we have derived
an expression for the amplitude expectation value that fully describes
the noisy supercritical bifurcation. The agreement with experiments
carried out for an 1D pattern forming system is excellent. This amplitude
equation can be applied to any second order transition of noisy temporal
systems.

We thank Rene Rojas for useful discussions. The simulation
software \textit{DimX} developed at the laboratory INLN in France
has been used for some numerical simulations. We thank the support
\emph{Anillo} grant ACT15 and the ECOS-Sud Committee. The CERLA is
supported in part by the ``Conseil R\'{e}gional Nord Pas de
Calais'' and the ''Fonds Europ\'{e}en de D\'{e}veloppement
Economique des R\'{e}gions''.

\end{document}